\documentclass[pre,aps,preprint,showpacs]{revtex4}
\usepackage{graphicx}
\usepackage{amsfonts}
\usepackage{amssymb}
\usepackage{amsmath}
\usepackage{textcomp}

\usepackage{color}
\usepackage{psfrag}

\usepackage{epsfig}

\begin{document}

\title{A Superior Descriptor of Random Textures and Its Predictive Capacity}

\author{Y. Jiao$^1$, F. H. Stillinger$^2$ and S. Torquato$^{2,3,4,5}$}


\affiliation{$^1$Department of Mechanical and Aerospace
Engineering, Princeton University, Princeton New Jersey 08544,
USA}



\affiliation{$^2$Department of Chemistry, Princeton University,
Princeton New Jersey 08544, USA}





\affiliation{$^3$Program in Applied and Computational Mathematics,
Princeton University, Princeton New Jersey 08544, USA}

\affiliation{$^4$School of Natural Sciences, Institute for
Advanced Study, Princeton NJ 08540}

\affiliation{$^5$Princeton Center for Theoretical Science,
Princeton University, Princeton New Jersey 08544, USA}

\begin{abstract}
 Two-phase random textures abound in a host of contexts, porous
and composite media, ecological structures, biological media and astrophysical structures.
Questions surrounding the spatial structure of such textures continue to
pose many theoretical challenges. For example, can  two-point correlation functions 
be identified that can be both
manageably measured and yet reflect nontrivial higher-order structural information
about the textures? We present a novel solution to this question
by probing the information content of the widest class
of different types of two-point  functions examined to date using
inverse ``reconstruction" techniques. This enables us
to show that a superior descriptor is the two-point cluster function
$C_2({\bf r})$, which is sensitive to topological {\it connectedness} information.
We demonstrate the utility of $C_2({\bf r})$ by accurately reconstructing 
textures drawn from materials science, cosmology and granular media,
among other examples. Our work suggests an entirely new theoretical pathway to 
predict the bulk physical properties of random textures, and also has important ramifications
for atomic and molecular systems.
\end{abstract}



\maketitle

Two-phase random textures  are ubiquitous in nature and synthetic
situations. Examples include heterogeneous materials (e.g.,
composites,  porous media and colloids) \cite{torquato,Sa03}, geologic media (e.g.,
rock formations) \cite{sandstone}, ecological structures (e.g., tree patterns in
forests) \cite{ecology}, cosmological structures (e.g., galaxy distributions) \cite{Pe93,Ga05}, 
and biological media (e.g., animal and plant tissue) \cite{Kh08}. 
Over a broad range of length scales, two-phase random textures
exhibit a rich variety of structures with varying degrees of disorder
and complex bulk properties \cite{To97,Zo06,Br07},
and questions concerning their quantitative characterizations continue
to present many fundamental and practical challenges.

It is well known that an infinite set of $n$-point correlation
functions is generally required to completely statistically characterize
such textures and their physical properties in the infinite-volume limit. A variety of different types
of correlation functions arise in rigorous theories of structure/property 
relations \cite{torquato}. One such basic quantity is the 
standard $n$-point correlation function $S_n({\bf x}_1,{\bf x}_2, \ldots, {\bf x}_n)$, 
which gives the probability of finding $n$ points at positions ${\bf x}_1,{\bf x}_2, \ldots, {\bf x}_n$
all in one of the phases \cite{torquato,To83}. Since the  information
contained in such an infinite set of functions
is generally unattainable in practice, a natural starting
point is to characterize the structure and bulk properties of
random textures using lower-order versions. The two-point
function $S_2$, which is experimentally accessible via scattering of radiation \cite{De49},
provides information about the distribution of pair separations. 
The three-point function $S_3$ reveals information about how these pair
separations involved in $S_2$ are linked into triangles. The four-point function $S_4$
controls the assembly of triangles into tetrahedra. A natural question is how much additional
useful information do $S_3$ and $S_4$ contain over and above $S_2$? 
We will show that triangular and tetrahedral statistics 
do not significantly increase information content over 
and above pair statistics for textures possessing no long-range order.

Therefore, an outstanding problem in condensed matter theory is to identify
other two-point correlation functions that can be both
manageably measured and yet reflect nontrivial higher-order structural information
about the textures. The aim of this paper is to provide
a novel solution to this problem using inverse techniques;
specifically, ``reconstruction" methods. The purpose of a reconstruction (construction) 
technique is to reconstruct (construct) realizations of two-phase
random textures that match limited structural information
on the textures in the form of lower-order correlation functions
\cite{torquato,Ye98a}, which are called the ``target" functions.
 An effective reconstruction procedure
enables one to generate accurate renditions of random textures at will \cite{Ye98b,Cu99,ApplyA,ApplyC,ApplyD,Ji07,Ji08}  and
subsequent analysis can be performed on the reconstruction to
obtain desired macroscopic properties of the texture nondestructively \cite{Ye98b,ApplyD}.
Here we use this inverse methodology to determine the amount of structural
information that is embodied in a set of targeted correlation function by quantifying
the extent to which the original structure can be accurately reconstructed
using those target functions. We quantify the accuracy of a reconstruction
by measuring {\it unconstrained} (untargeted) correlation
functions and comparing them to those of the original
medium.

We adapt the inverse reconstruction method of Yeong and Torquato \cite{Ye98a,Ye98b}
to show that a superior two-point signature of random textures is the two-point cluster function
$C_2({\bf r})$ \cite{To88}, which is sensitive to
topological {\it connectedness} information. We demonstrate that $C_2({\bf r})$
not only contains appreciably more information
than $S_2$, but more information than a variety of other ``two-point" quantities, including
the surface-surface correlation function $F_{ss}$, the
surface-void correlation function $F_{sv}$, the pore-size
function $F$, lineal-path function $L$,
and the chord-length  density function $p$
\cite{torquato}. (All the aforementioned correlation functions are defined precisely below.)
Such a comprehensive study that investigates and compares  all of the
aforementioned statistical descriptors via inverse techniques has heretofore not been
undertaken. Our results have practical implications for materials science, liquid-
and solid-state physics, biological systems, cosmology, hydrology and ecology.

\section{Definitions of the Two-Point Correlation Functions}

In the theory of random media, a variety of different types of  two-point correlation
functions naturally arise \cite{torquato}. Here we define seven different ones
that will be employed in this paper. 
Consider $d$-dimensional two-phase textures 
in which phase $i$ has volume fraction
$\phi_i$ ($i=1,~2$) and is characterized by the indicator function
\begin{equation}
\label{eq101} {\cal I}^{(i)}({\bf x}) = \left\{ {\begin{array}{*{20}c}
{1, \quad\quad {\bf x} \in {\cal V}_i,}\\
{\;\;0, \quad\quad \mbox{otherwise},}
\end{array} }\right.
\end{equation}
where ${\cal V}_i$ is the region occupied by phase $i$ (equal to 1 or 2). Note that 
here ``phase'' is used in a general sense in that it can refer to a  
solid, liquid or even void.
The standard two-point correlation function is defined as 
\begin{equation}
S^{(i)}_2({\bf x}_1,{\bf x}_2) = \left\langle{{\cal I}^{(i)}({\bf x}_1){\cal I}^{(i)}({\bf x}_2)}\right\rangle,
\end{equation}
where angular brackets denote an ensemble average.
This function is the probability
of finding two points ${\bf x}_1$ and ${\bf x}_2$  both in phase $i$. Henceforth,
we will drop the superscript ``$i$'' and only consider the
correlation functions for the phase of interest. 
For \textit{statistically homogeneous} and
\textit{isotropic} microstructures, which is the focus of the rest
of the paper, two-point correlation functions will
only depend on the distance $r \equiv |{\bf x}_1-{\bf x}_2|$ between the points
and hence $S_2({\bf x}_1,{\bf x}_2)=S_2(r)$.

The surface-void $F_{sv}$ and surface-surface  $F_{ss}$ correlation functions are 
respectively defined as 
\begin{equation}
F_{sv}(r) =
\left\langle{{\cal M}({\bf x}_1){\cal I}({\bf x}_2)}\right\rangle, \quad 
F_{ss}(r) = \left\langle{{\cal M}({\bf x}_1){\cal M}({\bf x}_2)}\right\rangle,
\end{equation}
where ${\cal M}({\bf x})=|\nabla {\cal I}({\bf x})|$ is the 
two-phase interface indicator function. By associating a
finite thickness with the interface, $F_{sv}$ and $F_{ss}$ can be
interpreted, respectively, as the the probability
of finding ${\bf x}_1$ in the ``dilated'' interface region and
${\bf x}_2$ in the void phase and the probability of finding both ${\bf x}_1$ and
${\bf x}_2$ in the ``dilated'' interface region but in the limit that the
thickness tends to zero \cite{torquato}. 

The lineal-path function $L(r)$ is the probability that 
an entire line of length $r$ lies in the phase of interest,
and thus contains a coarse level of {\it connectedness}
information, albeit only along a {\it lineal path} \cite{torquato,Lu92}.
The chord-length density function $p(r)$ gives the probability  
associated with finding a ``chord"  of length $r$
in the phase of interest and is directly proportional to the second derivative
of $L(r)$ \cite{To93}. (Chords are the line segments between the intersections of an
infinitely long line with the two-phase interface.)
The pore-size function $F(\delta)$ is related to the probability that
a sphere of radius $r$ can lie entirely in the phase of interest \cite{torquato}
and therefore is the three-dimensional ``spherical" version of the lineal measure $L$.

The two-point cluster function $C_2(r)$ gives the probability of finding two
points separated by a distance $r$ in the same {\it cluster} of the
phase of interest \cite{To88}. Note that a cluster of a phase is any topologically 
connected subset of that phase. The two-point cluster function
can be measured experimentally using any appropriate three-dimensional imaging technique
(e.g., tomography, confocal microscopy and MRI) \cite{torquato}.
The fact that $C_2$ contains intrinsic three-dimensional topological
information is to be contrasted with $S_2$, which can obtained
from a planar cross-section of the texture.
In general, $C_2$ is expected to embody a much greater level
of three-dimensional connectedness information than either $L$ or $F$, but the
degree to which this is true has yet to be quantitatively demonstrated,
which is one of the aims of this paper.

\section{Inverse Reconstruction Technique}

The stochastic optimization reconstruction algorithm for digitized media 
formulated  by Yeong and Torquato \cite{Ye98a}
is ideally suited to carry out the aforementioned analysis
because it can incorporate different types of target statistical descriptors. 
This algorithm is both robust and simple to implement \cite{Cu99,ApplyA,ApplyC,ApplyD,Ji07,Ji08}. 
In this method, one starts with an initial realization
of a random medium and a set of target correlation functions 
$\widehat{f}^{1}_n({\bf R})$, $\widehat{f}^{2}_n({\bf R})$,
$\widehat{f}^{3}_n({\bf R})$, $\ldots$, which are obtained
(i.e., measured) from the medium of interest. Here  ${f}^{\alpha}_n({\bf R})$
is an $n$-point correlation function of type $\alpha$, 
${\bf R} \equiv {\bf r}_1,{\bf r}_2, \ldots {\bf r}_n$,
and ${\bf r}_i$ denotes the position vector of the i{\em th} point.   The method proceeds to find a
realization in which calculated correlation functions
${f}^{1}_n({\bf R})$, ${f}^{2}_n({\bf R})$,
${f}^{3}_n({\bf R})$, $\ldots$ best match the corresponding target functions. This is
achieved by minimizing an ``energy" 
\begin{equation}
\label{eq208} E = \sum\limits_{\bf
R}\sum\limits_{\alpha}\left[{f^{\alpha}_n({\bf R})-\widehat{f}^{\alpha}_n({\bf R})}\right]^2,
\end{equation}
which is defined to be the sum of squared differences  between the calculated and target functions, 
via a simulated annealing method in which a sequence of trial
realizations is generated and accepted with the probability min$\{\exp(-\Delta E/T), 1\}$,
where $\Delta E$ is the energy difference between the new and old
realizations and $T$ is a fictitious temperature. 
The initially ``high" temperature is lowered according to a prescribed annealing schedule until the
energy of the system approaches its ground-state value within a very small 
tolerance level.

\subsection{Degeneracy of Ground States Using $S_2$ Alone}

It is instructive to consider the energy defined by (\ref{eq208})
when only the standard two-point correlation function $S_2$
is used. In this special instance, the energy is given by
\begin{equation}
E = \sum\limits_{r}\left[{S_2(r)-\widehat{S}_2(r)}\right]^2,
\label{energy_S2}
\end{equation} 
where $\widehat{S}_2$ and $S_2$ are the two-point correlation 
functions of the target and reconstructed medium, respectively.
We note that most previous reconstruction studies have only tried to match $S_2$.
However, it is now well established that
$S_2({\bf r})$ is not sufficient information to generally get an
accurate rendition of the original microstructure \cite{Ye98a,Ye98b,Cu99,Ji07,Ji08} 
In other words, 
the ground states when only $S_2$ is incorporated in the energy [Eq.~(\ref{energy_S2})] 
are highly degenerate due to the nonuniqueness of the 
information content of this two-point function, which is clearly 
illustrated by the subsequent examples in the paper. 
The reader is also referred to the {\it Supporting Text},
which provides a rigorous explanation for the degeneracy
of the ground states for digitized representations of textures.

\subsection{Insufficiency of Conventional Three-Point Information}

An obvious additional set of correlation functions that
could be incorporated in the reconstruction is the
higher-order versions of $S_2$, namely, $S_3$, $S_4$, etc. 
However, not only is the three-point
correlation function $S_3$ more difficult to compute, it is not at
all clear that its incorporation will result in appreciably better
reconstructions because it only introduces local information about
triangles when there is no long-range order, the most common
occurrence. We can quantitatively verify the insufficiency of
conventional triangular information by reconstructing a 
one-dimensional equilibrium distribution of equal-sized hard rods \cite{To36}
using $S_3$ (see the \textit{Supporting Text} for technical details). 
Figure \ref{fig_hardrod} compares this reconstruction to
those involving $S_2$ alone and a combination of $S_2$ and $C_2$.
It is clear that the reconstruction using $S_2$ only significantly
results erroneously in a highly clustered ``rod'' phase. 
Although incorporating $S_3$ provides an improved reconstruction, 
it still contains large clusters and isolated rods that can be much
smaller than actual rod size. On the other hand, the $S_2$-$C_2$ hybrid reconstruction 
produces the most accurate rendition of the target medium.
While this one-dimensional example is suggestive that $C_2$
contains nontrivial structural information in excess to what 
is contained in $S_2$, one must
investigate this problem in higher dimensions, which
presents algorithmic challenges, as we will describe 
in the next section.

\section{Efficient Algorithmic Implementation of the General Problem}

Here we present a general methodology that enables one to efficiently
incorporate a wide class of lower-order correlation functions
in the Yeong-Torquato reconstruction procedure. 
The aforementioned probabilistic interpretations of the correlation functions 
enable us here to develop a general sampling method for 
reconstruction of statistically
homogeneous and isotropic digitized textures based on 
the ``lattice-gas" formalism, which was introduced in Ref.~\cite{Ji08} and 
has been generalized here.
In the generalized formalism, pixels with different values (occupying the lattice sites) 
correspond to distinct local states
and pixels with the same value are considered to be
``molecules'' of the same ``gas" species \cite{Ji08}. 
The correlation functions
of interest can be obtained by binning the separation distances
between the selected pairs of molecules from particular species.

In the case of $S_2$, all ``molecules'' are of the same species. 
We denote the number of lattice-site separation distances of length $r$ by $N_S(r)$
and the number of molecule-pair separation distances of length $r$ by $N_P(r)$.
Thus, the fraction of pair distances with
both ends occupied by the phase of interest, i.e., the two-point
correlation function, is given by $S_2(r) = N_P(r)/N_S(r)$. To
obtain $C_2$, one needs to partition the ``molecules'' into
different subsets $\Gamma_i$ (``species'') such that any two
molecules of the same species are connected by a path composed of
the same kind of molecules, i.e., molecules that form a cluster,
which is identified using the ``burning'' algorithm
\cite{burning}. The number of pair distances of length $r$ between
the ``molecules'' within the same subset $\Gamma_i$ is denoted by
$N^i_P(r)$. The two-point cluster function is then given by
$C_2(r) = \sum_i N^i_P(r)/N_S(r)$. The calculation of $F_{ss}$ and
$F_{sv}$ requires partitioning the ``molecules'' into two
subsets: the surface set $\kappa_S$ containing only the
``molecules'' on the surfaces of the clusters and the volume set
$\kappa_V$ containing the rest. In a digitized medium, the
interface necessarily has a small but finite thickness determined
by the pixel size. Thus, the surface-surface and surface-void
correlation functions can be regarded as probabilities that are given by
$F_{ss} = N^{ss}(r)/N_S(r)$ and $F_{sv} = N^{sv}(r)/N_S(r)$,
respectively; where $N^{ss}(r)$ gives the number of distances
between two surface molecules with length $r$ and $N^{sv}$ is the
counterpart for pairs with one molecule on the surface and the
other inside the cluster. 

The lineal path function $L$ can be obtained by computing the 
lengths of all digitized line segments (chords) composed of pixels of the 
phase of interest, and for each chord incrementing the counters 
associated with the distances equal to and less than that chord length \cite{Ye98a,Ye98b}. 
The chord-length density function $p$ can then be easily obtained by binning 
the chord lengths that are used to compute $L$ \cite{torquato}. The pore-size function $F$ 
can be computed by finding the minimal separation distances of pixels within the phase 
of interest to those at the two-phase interfaces. The minimal distances are then 
binned to obtained a probability distribution function, the complementary cumulative 
distribution function of which is $F$ \cite{sandstone}.

We have also devised methods to
track clusters and surfaces that enable one to
quickly compute the correlation functions of the new realization
based on the old ones (see the \textit{Supporting Text} for 
technical details) and thus make the Yeong-Torquato reconstruction procedure much more
efficient than methods that directly re-sample the correlation functions for each
trial realization.

We have used this general procedure here to reconstruct a wide spectrum
of random textures, including model microstructures, such as the
cherry-pit model, equilibrium hard spheres,
Debye random media and symmetric
cell materials \cite{torquato}, as well as the digitized
representations of sandstones \cite{sandstone}, metal-ceramic
composites \cite{ceramic}, concrete microstructures \cite{concrete}, 
laser speckle patterns \cite{Ji08} and 
galaxy distributions. Using the aforementioned largest set
of correlation functions utilized to date, our analysis reveals
that the best reconstructions always incorporate the two-point
cluster function $C_2(r)$. In what follows, we will 
present specific results for only a subset of the correlation
functions that we used, namely, various combinations of $S_2$, $F_{ss}$, $F$ and $C_2$,
for the non-percolating phases of 
a concrete microstructure \cite{concrete}, a distribution of galaxies,
and a three-dimensional hard-sphere packing.

\subsection{Concrete Microstructure}

The wide range of structural features in concrete,  
from nanometer-sized pores to centimeter-sized aggregates, 
makes it a wonderful example of a multi-scale microstructure \cite{Ga98}.
Figure~\ref{fig-con}(a) shows a binarized digitized image of a concrete sample cross-section.
We have thresholded the original image so that 
the blue phase represents the stones  and the
lighter gray phase is the cement paste. The ``stone" phase 
is characterized by a dense dispersion of ``particles" of various sizes:
a nontrivial situation to reconstruct. Using $S_2$ alone overestimates 
clustering in the system and indeed incorrectly yields
a percolating ``particle'' phase. Thus, although $S_2$
of the reconstruction  matches the  target
one with very small error (see the figure in the \textit{Supporting Text}),  
such information is insufficient
to get a good reconstruction. Incorporating both $S_2$ and 
surface-surface function $F_{ss}$ leads to a better rendition of the target system but
the reconstruction still overestimates the degree of clustering.
On the other hand, incorporating
$C_2$ yields an excellent reconstruction
in that the ``stone'' phase clearly appears as a particle dispersion
with a size distribution that closely matches that of the target
structure. As noted in the introduction, we can quantitatively test
the  accuracy of the reconstructions by measuring unconstrained 
correlation functions and comparing them to the corresponding quantities
of the target system. Here we choose to compute the unconstrained
lineal-path function $L$. Figure~\ref{fig-con}(e) reveals
that the lineal-path function of the reconstruction that incorporates $C_2$ 
matches the target function $L$ well and it clearly is appreciably
more accurate than the other reconstructions.

\subsection{Galaxy Clusters}

Correlation functions have been used to understand
the formation of galaxies and the large scale structure of the Universe \cite{Pe93,Ga05}.
Such characterizations are becoming increasingly important
with the advent of high-quality surveys in cosmology.
We suggest that the reconstruction procedure
can provide an important tool in cosmological
and astrophysical applications, especially in 
helping to determine the lower-order
correlation functions that reflect {\it a priori} information
about nontrivial structural features, 
such as multi-scale clustering and filamentary structures
in the Universe. Figure~\ref{fig-galaxy}(a) shows a binarized image of a portion
of the Abell 1689 galaxy cluster. 
We have chosen the binarizing threshold such that the ``galaxy" phase (bright spots
in Fig.~\ref{fig-galaxy}a)
exhibits different sized clusters. 
As one can see, the ``primary'' (single largest) cluster in the target system has been
reproduced by all the reconstructions. This is because the volume fraction
of the  ``galaxy'' phase is relatively small while the target $S_2$ has a relatively
long tail (see the figure in the \textit{Supporting Text}), which requires the clustering of a majority of ``galaxy'' phase.
However, the ``secondary'' (smaller compact) clusters are significantly different for
different reconstructions. The reconstruction using $S_2$ only
produces large elongated secondary clusters. Incorporating $F_{ss}$ enables 
one to obtain a better rendition, however it still contains 
a few elongated secondary clusters. 
The incorporation of $C_2$ again provides the most accurate reconstruction. 
This is visually verified by examination of both the
sizes and shapes of the primary and secondary clusters, and is
quantitatively confirmed  by the comparison of the unconstrained $L$ of the 
reconstructed and target systems, as shown in Fig.~\ref{fig-galaxy}(e).

\subsection{Sphere Packing}

As an application of our methodology to three dimensions,
we have reconstructed a digitized realization of an equilibrium
distribution of equal-sized  hard spheres, 
as shown in Fig.~\ref{fig-sphere}(a). 
This packing is generated using the standard Metropolis Monte
Carlo technique for a canonical ensemble of 
hard spheres in a cubical box under periodic boundary 
conditions \cite{torquato}. A visual comparison of the hybrid reconstruction
involving the two-point cluster function reveals that it
accurately yields a dispersion of well-defined spherical inclusions
of the same size, in contrast to the $S_2$ reconstruction,
which again grossly overestimates clustering of the ``sphere" phase. 
In contrast to the previous examples, here we incorporate
the pore-size function $F$ (not the surface correlation
functions) in one of the reconstructions.
Although the reconstruction incorporating $F$ provides improvement over the 
rendition on the $S_2$-alone reconstruction, it is still inferior 
to the $S_2$-$C_2$ reconstruction in reproducing both the size and shape of the ``sphere'' phase.
(The target and reconstructed correlation functions are shown in 
the \textit{Supporting Text}.) The accuracy of the $S_2$-$C_2$ hybrid reconstruction 
can also be seen by comparing the unconstrained 
lineal-path function $L$ of the target system to those
of the reconstructed media [see Fig.~\ref{fig-sphere}(e)].

\section{Discussion}

In summary, while it was known that the information content 
of the standard two-point function $S_2$ of a random texture is far from complete, 
we have demonstrated here that the next higher-order version $S_3$
generally does not contain appreciably greater information. The
fact that this natural extension to incorporate higher-order $S_n$,
which has been pursued in the last century in statistical mechanics,
is not a fruitful path motivated us to inquire whether there exist
sensitive two-point statistical descriptors  that embody nontrivial
structural information. We probed the information content
of seven different types of two-point functions
using inverse reconstruction methods.
For all of the examples studied here,
reconstructions that include the two-point cluster function $C_2$
were always found to be significantly more accurate
than those involving any of the combinations of pairs of the other functions.

\begin{figure}[bthp]
\begin{center}
\includegraphics[width=9.0cm,keepaspectratio,clip=]{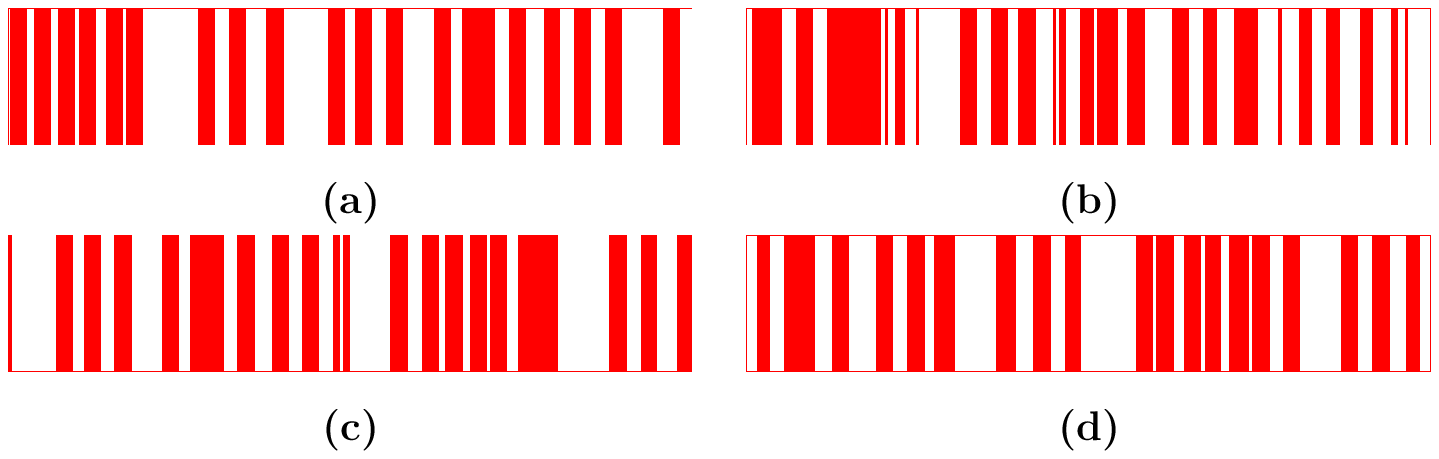} \\
\end{center}
\caption{(a) Target: an equilibrium hard-rod system in which the equal-sized rods cover 50\% space. 
Each rod in the system is 10 pixels in length. (b) $S_2$-alone reconstruction. 
(c) $S_3$ reconstruction. (d) $S_2$-$C_2$ hybrid reconstruction. 
For visualization purposes, the one-dimensional rod systems are
artificially extended in the vertical direction.}
\label{fig_hardrod}
\end{figure}

\begin{figure}[bthp]
\begin{center}
\includegraphics[width=7.5cm,keepaspectratio]{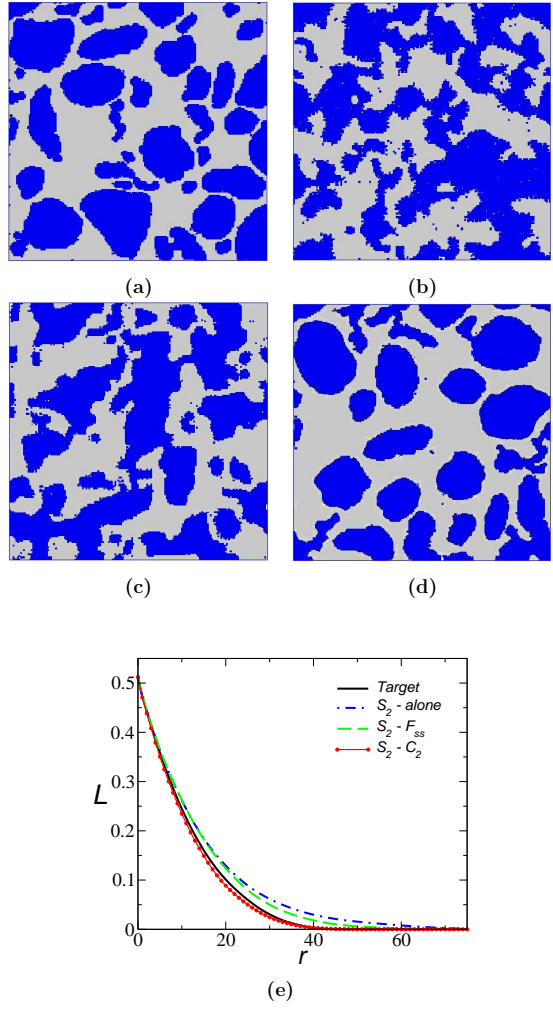} \\
\end{center}
\caption{(a) Target system: a binarized image
 of a cross-section of concrete \cite{concrete}. The linear size 
of the digitized texture is $N_L=170$ pixels.
(b) $S_2$-alone reconstruction. (c) $S_2$-$F_{ss}$ hybrid
reconstruction. (d) $S_2$-$C_2$ hybrid reconstruction. 
All the reconstructions are associated with a final energy (error) $E \sim 10^{-8}$.
(e) The unconstrained lineal-path function $L$ of the reconstructions and the target image. 
Pixel size supplies the unit for the distance $r$.}
\label{fig-con}
\end{figure}

\begin{figure}[bthp]
\begin{center}
\includegraphics[width=7.5cm,keepaspectratio]{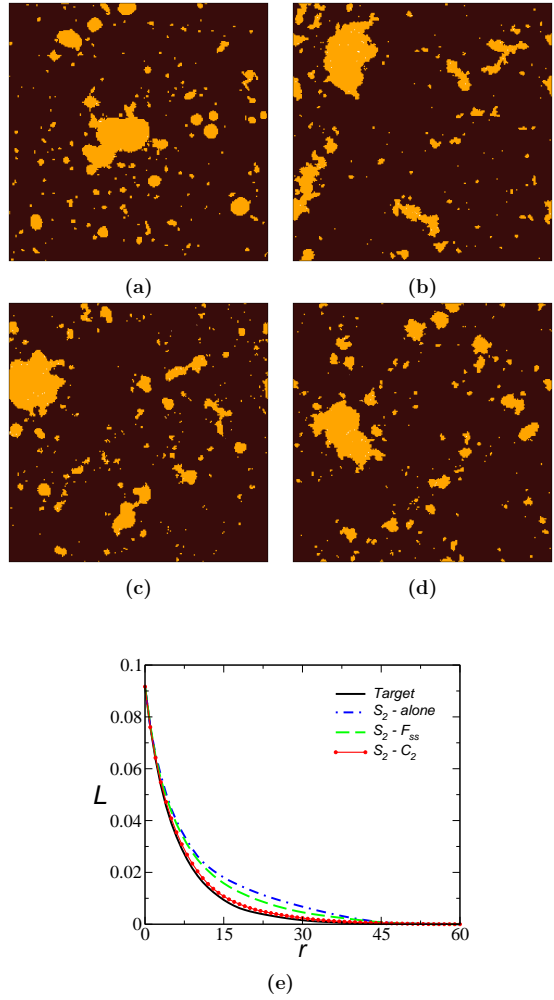} \\
\end{center}
\caption{(a) Target system: a portion of the Coma Cluster. 
The linear size of the digitized texture is $N_L=238$ pixels.
(b) $S_2$-alone reconstruction. (c) $S_2$-$F_{ss}$ hybrid
reconstruction. (d) $S_2$-$C_2$ hybrid reconstruction.
All the reconstructions are associated with a final energy (error) $E \sim 10^{-8}$.
(e) The unconstrained lineal-path function $L$ of the reconstructed and target systems. 
Pixel size supplies the unit for the distance $r$.}
\label{fig-galaxy}
\end{figure}

\begin{figure}[bthp]
\begin{center}
\includegraphics[width=7.5cm,keepaspectratio]{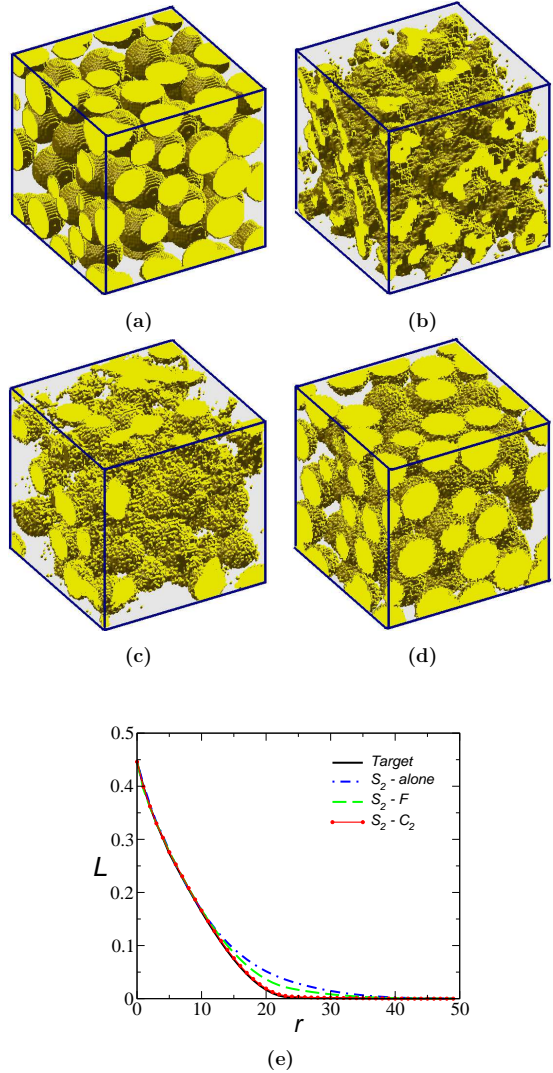} \\
\end{center}
\caption{(a) Target system: a digitized realization of a
hard-sphere packing in which the spheres
occupy 44.6\% of space. The linear size 
of the digitized texture is $N_L=100$ pixels.
(b) $S_2$-alone reconstruction. (c) $S_2$-$F$ reconstruction.
(d) $S_2$-$C_2$ hybrid reconstruction.
All the reconstructions are associated with a final energy (error) $E \sim 10^{-11}$.
(e) The (unconstrained) lineal-path function sampled from target and reconstructed realizations. 
Pixel size supplies the unit for the distance $r$.}
\label{fig-sphere}
\end{figure}

\begin{figure}[bthp]
\begin{center}
\includegraphics[width=7.5cm,keepaspectratio]{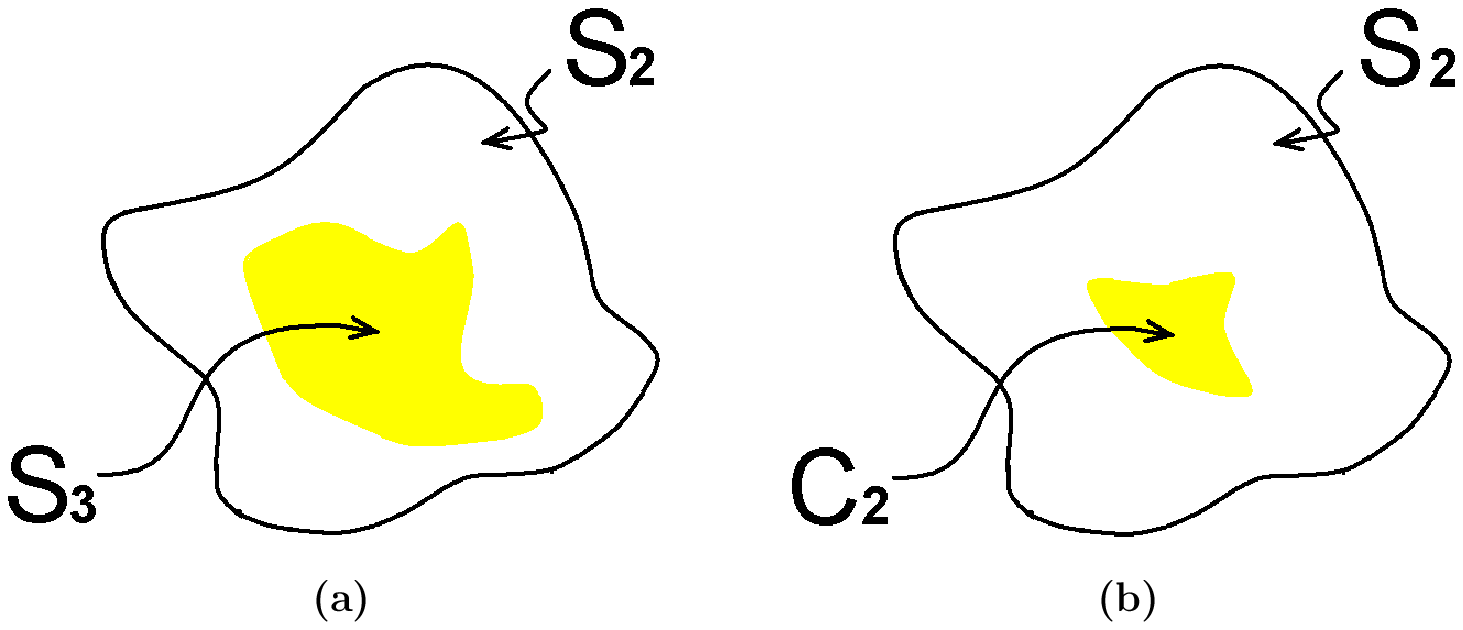} \\
\end{center}
\caption{The set of all microstructures associated with a
particular $S_2$ is schematically
shown as the region enclosed by the solid contour in both panels.
The shaded region in the left panel shows the set of all
microstructures associated with the same $S_2$ and $S_3$. The shaded
more restrictive region in the right panel shows the set of all microstructures
associated with the same $S_2$ and $C_2$.} \label{fig00}
\end{figure}

More precisely, the incorporation of $C_2$ significantly reduces the
number of compatible microstructures as compared to the compatible
microstructures consistent with the same three-point function $S_3$, 
which is schematically indicated
in Fig.~\ref{fig00}. The two-point cluster function is an especially sensitive
structural signature  when clustering and phase connectedness
are present, the most difficult situations to treat. 
This can be seen from the reconstruction of the hard-rod system 
that incorporated $C_2$ (see Fig.~\ref{fig_hardrod}).
More importantly, we showed the utility of $C_2({\bf r})$ in higher
dimensions by accurately reconstructing
galaxy distributions, concrete microstructures and dense hard-sphere packings,
among other examples.

Why is $C_2(r)$ a superior two-point structural signature? To answer this question,
it is useful to first compare it to $S_2(r)$. The latter, unlike $C_2(r)$,
does not distinguish between events in which the end points of the line segment
of length $r$ fall in the same cluster of a particular phase
and those that do not involve the same cluster of that phase \cite{To88}. More precisely,
$C_2$ is the ``connectedness"
contribution to the standard two-point correlation function, i.e.,
\begin{equation}
S_2(r)=C_2(r)+D_2(r),
\end{equation}
where $D_2$ measures the probability that the end points of a line segment
of length $r$ fall in different clusters of the phase of interest.
Therefore,  whereas $S_2(r)$ is insensitive to clustering and percolation,
$C_2(r)$ becomes a progressively longer-ranged function as clusters grow
in size such that its volume integral
diverges at the percolation threshold \cite{To88,torquato}.
By contrast, the quantities $L$, $p$, $F$, $F_{ss}$ and $F_{sv}$ are insensitive
to crossing the percolation threshold. Indeed, for particle
systems, one can show that $C_2$ is a functional of the infinite
set of ``$n$-particle" connectedness functions \cite{To88}.

Thus, although $C_2$ is a ``two-point" quantity, it 
actually embodies higher-order structural
information in a way that makes it a highly sensitive
statistical descriptor over and above $S_2$. This ability
to ``leap frog" past the usual approach of incorporating
additional information via higher-order versions of
$S_2$ has important ramifications for new structure/property
of random textures. Specifically, our work suggests
that new theories should be developed that relate the transport, mechanical, chemical
and optical properties of random textures to functionals
that incorporate $C_2$. It is
clear that such theories would be highly predictive, since
it is well established that the presence of clusters in
textures can dramatically alter their macroscopic
physical properties \cite{torquato}.

It should not go unnoticed that our work also has important implications
for atomic and molecular systems where the analogous standard correlation
functions that arise are the two-body correlation function $g_2$, three-body
correlation function $g_3$, etc. \cite{Wi66,Ri66,He67,Ha06} The three-body function $g_3$, for example,
has been the focus of great attention, e.g., integral equations and approximations
have been devised for $g_3$ \cite{Ri66,He67,Ha06}. The spatial structure of disordered atomic or molecular systems
may be regarded to be special cases of random textures. For example, the particles that comprise
simple atomic systems are fully specified by their center-of-mass coordinates.
These point distributions can be decorated in an infinite number of ways
to produce random textures in the sense that we have defined them in this paper.
For example, one could circumscribe the points by spheres of  size 
dictated by the physics of the problem (e.g., electron hopping distance). For such systems, our work
suggests that the pair-connectedness function $P_2$ (i.e., the 
connectedness contribution to $g_2$ \cite{torquato}) should
contain far greater information than $g_3$ beyond that contained in $g_2$.

\begin{acknowledgments}
This work was supported by the American Chemical Society Petroleum Research Fund and the
Office of Basic Energy Sciences, U.S. Department of Energy,
under Grant No. DE-FG02-04-ER46108.
\end{acknowledgments}

\newpage

\begin{thebibliography}{10}
\bibitem{torquato}
Torquato S (2002) \textit{Random Heterogeneous Materials:
Microstructure and Macroscopic Properties} (Springer-Verlag, New
York).

\bibitem{Sa03}
Sahimi M (2003) {\it Heterogeneous Materials} (Springer-Verlag, New York).


\bibitem{sandstone}
Coker DA, Torquato S, Dunsmuir J (1996) Morphological and physical properties 
of Fountainebleau sandstone from tomographic analysis. J Geophys Res {101}: 17497-17510.


\bibitem{Pe93}
Peebles PJE (1993) {\it Principles of Physical Cosmology}
(Princeton University Press, Princeton, NJ).

\bibitem{Ga05}
Gabrielli A, Sylos Labini F, Joyce M, Pietronero P (2005)
\textit{Statistical Physics for Cosmic Structures}
(Springer-Verlag, New York).

\bibitem{ecology}
Pommerening A, Stoyan D (2008) Edge-correction needs in estimating 
indices of spatial forest structure. Can J For Res {38}: 1110-1122.

\bibitem{Kh08}
Kherlopian AR, Song T, Duan Q, Neimark MA, Po MJ, Gohagan JK, Laine AF (2008) 
A review of imaging techniques for systems biology. BMC Syst Biol {2}: 1-18.


\bibitem{To97}
Torquato S (1997) Exact expression for the effective elastic tensor of 
disordered composites. Phys Rev Lett {79}: 681-684.


\bibitem{Zo06}
Zohdi TI (2006) On the optical thickness of disordered particulate media. Mech Mater {38}: 969-981.


\bibitem{Br07}
Mejdoubi A, Brosseau C (2007) Numerical calculations of the intrinsic 
electrostatic resonances of artificial dielectric heterostructures. J Appl Phys {101}: 084109.

\bibitem{To83}
Torquato, S and Stell, G (1983) Microstructure of two-phase random media. II. 
The Mayer-Montroll and Kirkwood-Salsburg jierarchies,  J Chem Phys {78}: 3262-3272.


\bibitem{De49}
Debye P, Bueche AM (1949) Scattering by an inhomogeneous solid. J Appl Phys {20}: 518-525.


\bibitem{Ye98a}
Yeong CLY, Torquato S (1998) Reconstructing random media. Phys Rev E {57}: 495-506.

\bibitem{Ye98b}
Yeong CLY, Torquato S (1998) Reconstructing random media: II. 
Three-dimensional media from two-dimensional cuts. Phys Rev E {58}: 224-233.

\bibitem{Cu99}
Sheehan N, Torquato S (2001) Generating microstructures with 
specified correlation functions. J Appl Phys {89}: 53-60.

\bibitem{ApplyA}
Wu K, Dijke MIJ, Couples GD, Jiang Z, Ma J, Sorbie KS, Crawford J, Young I,  Zhang X (2006) 
3D stochastic modelling of heterogeneous porous media - applications to reservoir rocks. {Trans Porous Media}
{65}: 443-467.


\bibitem{ApplyC}
Basanta D, Miodownik MA, Holm EA, Bentley PJ (2005)
Investigating the evolvability of biologically inspired CA. {Metall Mater Trans A} {36}: 1643-1652.

\bibitem{ApplyD}
Kumar H, Briant CL, Curtin WA (2006) Using microstructure reconstruction to 
model mechanical behavior in complex microstructures. Mech Mater {38}: 818-832.


\bibitem{Ji07}
Jiao Y, Stillinger FH, Torquato S (2007) Modeling heterogeneous materials via 
two-point correlation functions: Basic principles. Phys Rev E {76}: 031110.

\bibitem{Ji08}
Jiao Y, Stillinger FH, Torquato S (2008) Modeling heterogeneous materials via 
two-point correlation functions. II. Algorithmic details and applications. Phys Rev E {77}: 031135.

\bibitem{To88}
Torquato S, Beasley JD, Chiew YC (1988) Two-point cluster function for 
continuum percolation. J Chem Phys {88}: 6540-6546.


\bibitem{Lu92}
Lu B, Torquato S (1992) Lineal-path function for random heterogeneous materials. 
Phys Rev A {45}: 922-929. 


\bibitem{To93}
Torquato S, Lu B (1993) Chord-length distribution function for 
two-phase random media.  Phys Rev E  {47}: 2950-2953. 

\bibitem{To36}
Tonks L (1936) The complete equation of state of one, two and three dimensional 
gases of hard elastic spheres. Phys Rev {50}: 955963.



\bibitem{burning}
Stauffer D, Aharony A (1994) \textit{Introduction to Percolation Theory} (Taylor \& Francis, London).




\bibitem{ceramic}
Torquato S, Yeong CLY, Rintoul MD, Milius DL, Aksay IA (1999) 
Characterizing the structure and mechanical properties of interpenetrating 
multiphase cermets. J Am Ceram Soc {82}: 1263-1268.


\bibitem{concrete}
Askeland DR, Phule PP (2005) \textit{ The Science and Engineering of Materials} 
(Cengage,  Florence, KY).



\bibitem{Ga98}
Garboczi EJ, Bentz DP (1998) Multi-scale analytical/numerical theory of the 
diffusivity of concrete. J Adv Cement-Based Mater {8}: 77-88.

\bibitem{Wi66}
Widom B (1968) Random Sequential Addition of Hard Spheres to a Volume. J Chem Phys 44: 3888-3894. 

\bibitem{He67}
Henderson D (1967) Structure of the Triplet Distribution Function. J Chem Phys 46: 4306-4310.

\bibitem{Ri66}
Rice SA, Gray P (1966) {\it The Statistical Mechanics of Simple Liquids} (John \& Wiley, New York).

\bibitem{Ha06}
Hansen JP, McDonald IR (2006) {\it Theory of Simple of Liquids} (Academic, New York).


\end{thebibliography}
\end{document}